\documentclass[aps, prb, reprint]{revtex4-2} 
\usepackage{fix-cm}
\usepackage[T1]{fontenc}

\usepackage{amsmath}  
\usepackage{amsfonts} 
\usepackage{graphicx} 
\usepackage{dcolumn}
\usepackage{bm}
\usepackage{xcolor}
\usepackage{float}
\usepackage{subcaption}

\usepackage{titlesec}
\titlespacing*{\section}{0pt}{\baselineskip}{\baselineskip}
\usepackage[utf8]{inputenc}
\usepackage[T1]{fontenc}
\usepackage{mathptmx}
\usepackage{etoolbox}
\usepackage{caption}
\begin{document}



\title{Visualizing Mathieu-Type Dynamics in a Tabletop Magnetic Trap: A Coil-Driven Parametric Oscillator}




\author{William Ho}
\email{williamho@college.harvard.edu}
\author{Anna Klales}
\email{aklales@g.harvard.edu}
\author{Daniel Davis} 
\email{ddavis@fas.harvard.edu}
\author{Jieping Fan}
\email{jieping.fang@gmail.com}
\author{Robert Hart}
\email{roberthart56@gmail.com}
\author{Ali Kurmus}
\email{akurmus@uw.edu}
\author{Louis Deslauriers}
\email{louis@physics.harvard.edu}
\affiliation{Department of Physics, Harvard University, Cambridge, MA 02138}














\date{\today}

\begin{abstract}
We present a tabletop demonstration of dynamic stabilization and ponderomotive-like trapping using a pair of sinusoidally-driven anti-Helmholtz coils and a suspended permanent magnet. The oscillating field produces a rapid micromotion superimposed on a slower secular oscillation, with micromotion amplitude increasing with displacement and peaking near the turning points. This behavior reveals a ponderomotive-like mechanism: a spatial gradient of micromotion amplitude that drives slow secular motion. The time-averaged effect provides a time-averaged harmonic (ponderomotive) restoring force that confines the magnet between the coils. Driving at 12--18\,Hz places the system in a small-$q_{\rm eff}$ regime where the two time scales are clearly separated and directly visible to the eye. Video tracking (included with this article) quantifies the motion and reveals a stability edge as the drive frequency is lowered (near 6--7\,Hz in our apparatus).
From trajectories in the 12--18\,Hz range, we extract an effective Mathieu parameter $q_{\rm eff}\approx 0.16$ from the measured timescale separation $\omega_{\rm sec}/\Omega$. The apparatus uses inexpensive, readily available parts, and we provide a concise materials list, analysis code, field-gradient calibration data, and demonstration videos.
\end{abstract}

\maketitle 

\section{INTRODUCTION}\label{sec:intro}

A system that is unstable under static conditions can become stable when subjected to an oscillating drive, a phenomenon known as dynamic stabilization. This principle appears in many areas of physics from Kapitza’s inverted pendulum to RF Paul traps, stabilized Rydberg states, and astrophysical rotating saddles near charged binary black holes \cite{Kapitza_1951,Paul_1990, Leibfried_2003,Bialynicki-Birula_1994,Kurmus_2025v2}. 
A widely-used classroom illustration to display dynamic stability is the rotating saddle, where a steel ball can be confined on a saddle rotating about its center point \cite{Rueckner_1995,Kirillov_2016,Fan_2017, Thompson_2002,lofgren_2023}. It achieves confinement through rigid rotation, and its trajectories are analytically solvable; however, rolling and contact friction can obscure the rapid micromotion which produces the effective restoring force and is the key signature of parametrically stabilized motion.

By contrast, many parametrically stabilized systems are driven by an oscillation of curvature at fixed orientation (“flapping” modulation) rather than by rigid rotation \cite{Thompson_2002}. In RF Paul traps, this is realized by a sinusoidal quadrupole that alternates between focusing and defocusing. The resulting equation of motion has periodic coefficients (Mathieu form; $y''(\tau)+[a-2q\cos(2\tau)]y(\tau)=0$, where $a$ and $q$ parametrize the static and oscillatory curvature), for which Floquet methods provide the standard analysis \cite{Paul_1990, Leibfried_2003, magnus_1966}. In the small‑\(q\) regime relevant here, the motion separates into two visible time scales: a slow \emph{secular} oscillation and a fast, phase‑locked \emph{micromotion} at the drive frequency, with micromotion amplitude increasing with distance from the center \cite{Wineland_1998}.


In this tabletop demonstration, we present Mathieu-type dynamic stabilization using a pair of sinusoidally-driven anti-Helmholtz coils and a suspended ring magnet. Suspending the magnet on a long, low-mass string eliminates contact friction from any guide surface while introducing only a weak, positive static curvature (\(a>0\)) set by the pendulum frequency \cite{carey_2025}. With drive frequencies chosen so that the two time scales are visible to the human eye, the trajectories exhibit micromotion that is largest near the secular turning points and smallest near the trap center. 
We observe a stability edge as the drive is lowered, and we extract an effective Mathieu parameter $q_{\rm eff}$ from 12–18\,Hz runs. We also provide videos of the motion, a list of critical components, analysis code, and a calibration of magnetic field gradient versus current.

\section{BACKGROUND THEORY}\label{sec:rf_vs_rotating}

\subsection{On-axis field of an anti-Helmholtz pair}

The magnetic field along the common symmetry axis ($x$-axis) produced by two coaxial coils
of radius $R$, separated by distance $d$, carrying equal currents in opposite directions
(anti-Helmholtz configuration; Fig.~\ref{fig:antihelmholtz}) is \cite{Youk_2005}
\begin{equation}
B_x(x) = \frac{\mu_0 I R^2}{2}\left[\frac{1}{\left(R^2 + \left(x-\frac{d}{2}\right)^2\right)^{3/2}} \;-\;
\frac{1}{\left(R^2 + \left(x+\frac{d}{2}\right)^2\right)^{3/2}}\right].
\label{eq:Bx_exact}
\end{equation}
This expression is odd about the midpoint, $B_x(-x)=-B_x(x)$, so its Taylor series about
$x=0$ contains only odd powers,
\begin{equation}
B_x(x) = G\,x + C\,x^3 + O(x^5).
\label{eq:Bx_taylor}
\end{equation}
The leading coefficient is
\begin{equation}
G = \left.\frac{dB_x}{dx}\right|_{x=0}
= \frac{3\mu_0 I R^2 d}{2\left(R^2+\frac{d^2}{4}\right)^{5/2}},
\label{eq:G_def}
\end{equation}
which yields the familiar linear approximation near the trap center ($|x|\ll R,d$).
Retaining the next nonzero term gives
\begin{equation}
C=\frac{1}{6}\left.\frac{d^3B_x}{dx^3}\right|_{x=0}
=
\frac{5\mu_0 I R^2 d\left(d^2-3R^2\right)}{4\left(R^2+\frac{d^2}{4}\right)^{9/2}}.
\label{eq:C_def}
\end{equation}
In what follows, we take $I$ in Eqs.~\eqref{eq:Bx_exact}--\eqref{eq:C_def} to denote the peak drive current $I_0$.

For a sinusoidal drive current with frequency $\Omega$,
\begin{equation}
I(t)=I_0\cos(\Omega t),
\end{equation}
the on-axis field factorizes as $B_x(x,t)=\cos(\Omega t)\,B_x(x;I_0)$. Using the midpoint
expansion in Eq.~\eqref{eq:Bx_taylor}, we write
\begin{equation}
B_x(x,t)=\cos(\Omega t)\,\big(Gx+Cx^3+O(x^5)\big),
\end{equation}
so the on-axis gradient is even in $x$,
\begin{equation}
\frac{\partial B_x}{\partial x}(x,t)=\cos(\Omega t)\,\big(G+3Cx^2+O(x^4)\big),
\label{eq:bprime_expand}
\end{equation}
and is generally nonzero at $x=0$.

\subsection{Equation of motion and micromotion envelope}

Approximating the suspended magnet as a dipole of moment $\mu$ aligned with the coil axis,
the axial force is
\begin{equation}
F_x=\mu\,\frac{\partial B_x}{\partial x}.
\end{equation}
Including weak static restoring curvature from the suspension ($k_p$) and linear damping
($\gamma$), the axial equation of motion is
\begin{equation}
m\ddot{x}+\gamma\dot{x}+k_p x
= \mu\,\frac{\partial B_x}{\partial x}(x,t).
\label{eq:eom_general}
\end{equation}
Using Eq.~\eqref{eq:bprime_expand} and keeping the leading even terms gives, near the midpoint,
\begin{equation}
m\ddot{x}+\gamma\dot{x}+k_p x
=
\cos(\Omega t)\Big(F_c+\alpha x^2+O(x^4)\Big),
\label{eq:eom_midpoint}
\end{equation}
where $F_c\equiv \mu G$ and $\alpha\equiv 3\mu C$ \cite{magnus_1966, nayfeh_1979}.

In the high-frequency regime ($\omega_{\rm sec}\ll\Omega$), we write the motion as
$x(t)=X(t)+\xi(t)$, where $X(t)$ varies slowly (secular motion) and $\xi(t)$ is a rapid
oscillation at the drive frequency \cite{dehmelt_1968}. Treating $X(t)$ as approximately constant over a few
drive cycles and keeping the leading-order fast response, Eq.~\eqref{eq:eom_midpoint} gives
\begin{equation}
\xi(t)\simeq -\frac{F_c+\alpha X^2}{m\Omega^2}\cos(\Omega t),
\end{equation}
so the micromotion envelope amplitude is
\begin{equation}
A_{\rm mm}(X)\;\simeq\;\frac{F_c+\alpha X^2}{m\Omega^2}
\;\equiv\;
A_{\rm floor}+\beta X^2,
\label{eq:Am_floor_beta}
\end{equation}
with
\begin{equation}
A_{\rm floor}\equiv \frac{F_c}{m\Omega^2},
\qquad
\beta\equiv \frac{\alpha}{m\Omega^2}.
\end{equation}
Equation~\eqref{eq:Am_floor_beta} predicts that micromotion is minimized but generally nonzero
at the center ($A_{\rm floor}$) and increases symmetrically away from the midpoint (leading
term $\propto X^2$). A useful consistency check is the ratio
\begin{equation}
\frac{A_{\rm mm}(X)}{A_{\rm floor}}
\;\approx\;
\frac{\left.\partial_x B_x\right|_{x=X}}{\left.\partial_x B_x\right|_{x=0}}
\;\approx\;
1+\frac{\alpha}{F_c}X^2
\;=\;
1+\frac{3C}{G}X^2,
\label{eq:mm_ratio_check}
\end{equation}
showing explicitly that the growth of micromotion away from the center is governed by the
dimensionless combination $(3C/G)X^2$ rather than by $C$ or $G$ separately.

\subsection{Ponderomotive restoring force}

The rapid oscillatory forcing in RHS of Eq.~\eqref{eq:eom_midpoint} has an amplitude
\begin{equation}
F_0(x)=F_c+\alpha x^2+O(x^4),
\end{equation}
which produces the standard time-averaged (ponderomotive) potential
\begin{equation}
U_{\rm pond}(x)=\frac{F_0^2(x)}{4m\Omega^2},
\label{eq:Up_def}
\end{equation}
derived by averaging over the fast drive cycle (see Appendix~\ref{appendix:ponderomotive_derivation}).
Expanding about $x=0$ yields
\begin{equation}
U_{\rm pond}(x)\simeq \text{const}+\frac{F_c\alpha}{2m\Omega^2}\,x^2+O(x^4),
\end{equation}
so the leading ponderomotive restoring force is linear in displacement even though the
micromotion envelope is not. The corresponding effective spring constant and secular frequency are
\begin{equation}
k_{\rm eff}=\frac{F_c\alpha}{m\Omega^2},
\qquad
\omega_{\rm sec}^2\simeq \frac{k_p}{m}+\frac{F_c\alpha}{m^2\Omega^2}.
\label{eq:omega_sec}
\end{equation}

\subsection{Secular--micromotion separation in the high-frequency regime}

We parameterize the separation between secular motion and micromotion by the dimensionless
quantity
\begin{equation}
q_{\rm eff}\equiv 2\sqrt{2}\,\frac{\omega_{\rm sec}}{\Omega},
\label{eq:qeff_def}
\end{equation}
so that $q_{\rm eff}\ll 1$ corresponds to $\omega_{\rm sec}\ll\Omega$. This definition is chosen
to match the conventional Mathieu $q$ parameter used for RF Paul traps, where (for $a\simeq 0$)
the secular frequency satisfies $\omega_{\rm sec}\approx (q/2\sqrt{2})\,\Omega$ in the small-$q$
regime \cite{Wineland_1998,Paul_1990}. We adopt $q_{\rm eff}$ as a convenient way to express the
same timescale separation in this mechanical analog and to facilitate comparison with rotating-saddle
analogs \cite{Thompson_2002}.

In the anti-Helmholtz geometry, the on-axis field is odd about the midpoint and the on-axis
gradient is even; consequently the micromotion is minimized but generally nonzero at the center
and increases symmetrically with displacement, as summarized by Eq.~\eqref{eq:Am_floor_beta}.
This position-dependent micromotion is the physical origin of the ponderomotive restoring
mechanism derived above.

\begin{figure}[h!]
    \centering
    \includegraphics[width=0.35\textwidth]{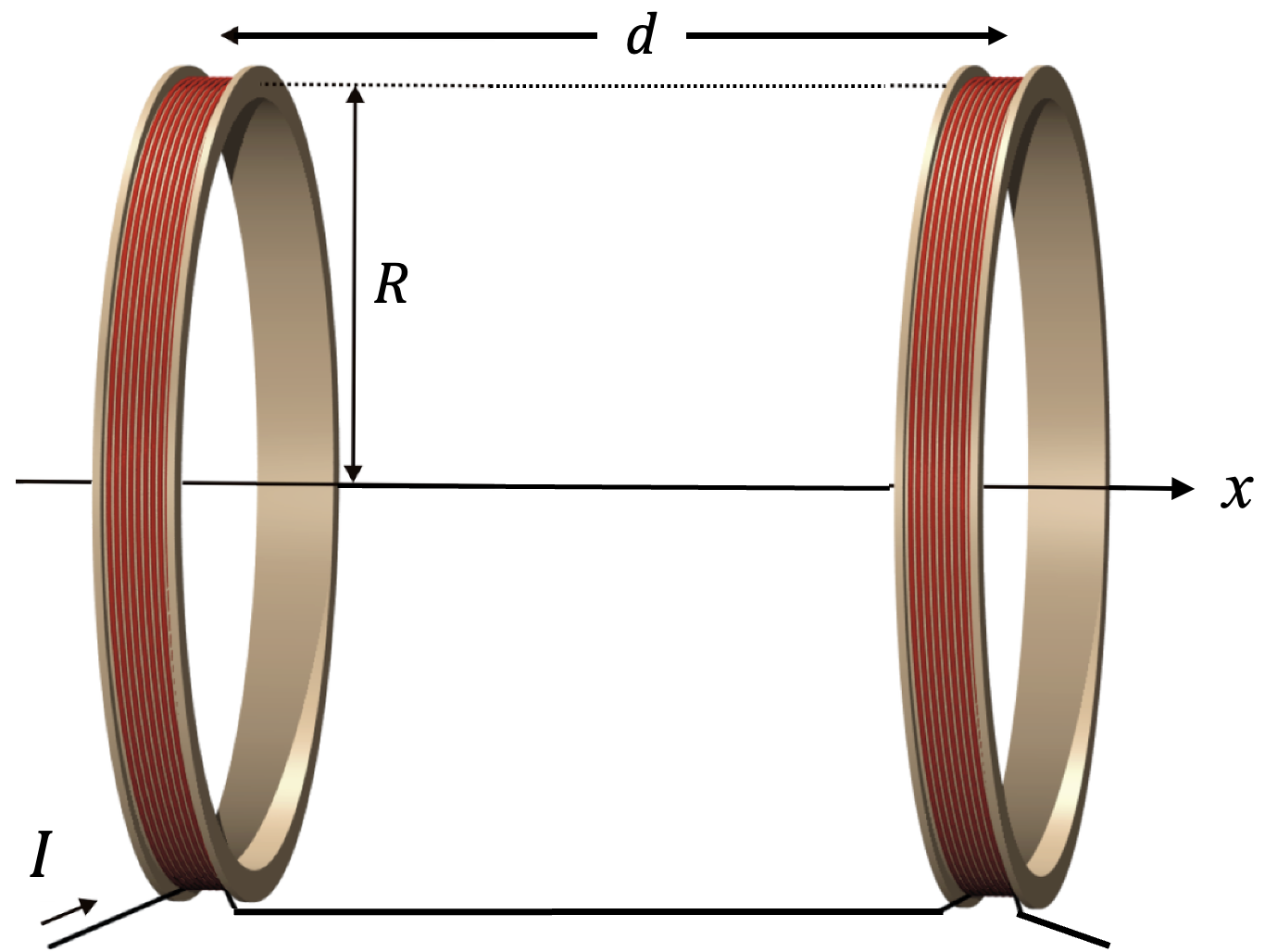}   
    \caption{Geometry of the anti-Helmholtz coil pair used to generate a magnetic field with a linear axial gradient around a central null point. The two coils of radius $R$ are separated by distance $d$ and carry equal currents in opposite directions. Although the full magnetic field is quadrupolar, only the axial component of its time-varying gradient is relevant in this one-dimensional demonstration.}
    \label{fig:antihelmholtz}
\end{figure}

\section{EXPERIMENTAL SETUP}
\label{sec:exp_setup}
\begin{figure}[h!]
    \centering
    \includegraphics[width=0.4\textwidth]{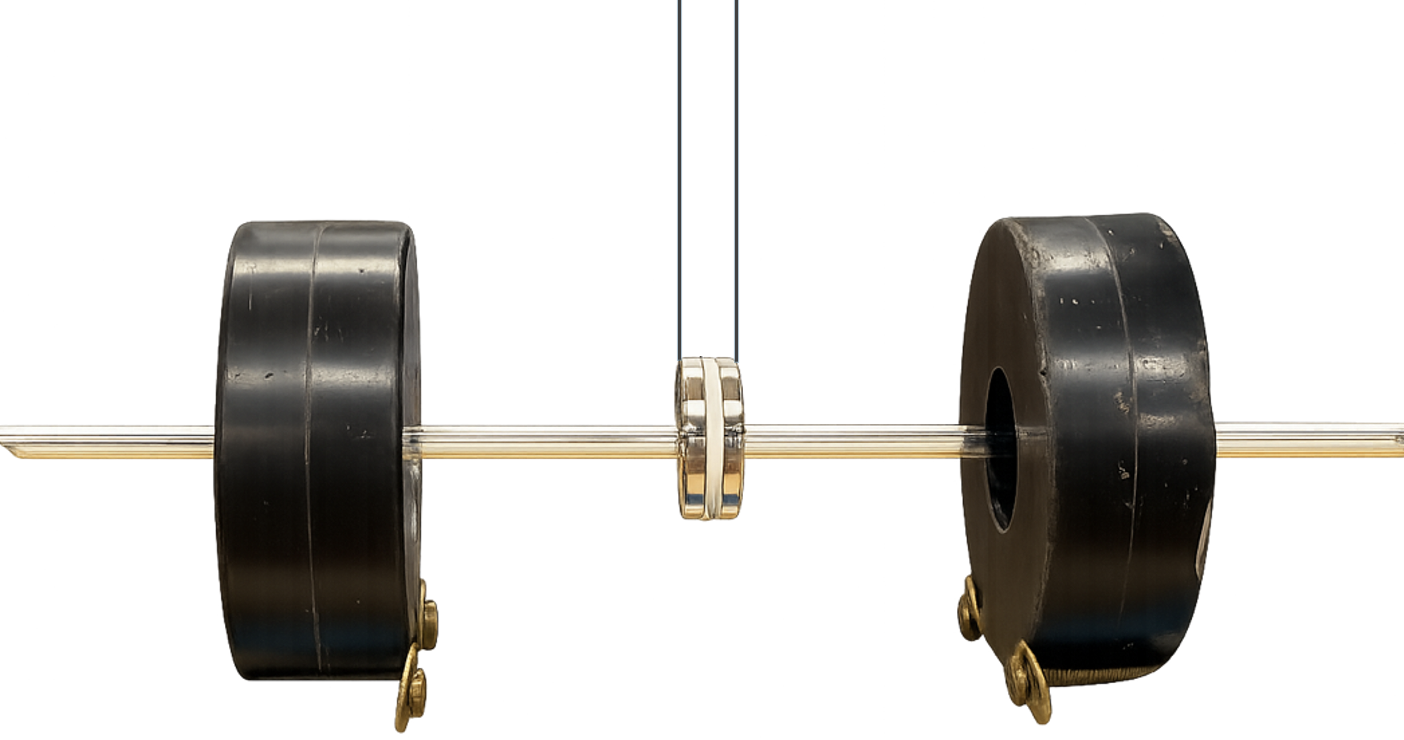}   
    \caption{Photograph of the experimental apparatus. A donut-shaped NdFeB ring magnet is suspended from a long thread between a coaxial anti-Helmholtz coil pair. The coils carry equal currents in opposite directions, producing a central field null and an approximately linear axial field gradient near the center.}
    \label{fig:Apparatus}
\end{figure}

Figure~\ref{fig:Apparatus} shows the apparatus used in this demonstration. The coil geometry
(radius $R$ and separation $d$) used in Sec.~II is defined in Fig.~\ref{fig:antihelmholtz}.
Two anti-Helmholtz coils (6.8\,cm OD $\times$ 2.3\,cm ID $\times$ 2.4\,cm thick) are mounted
coaxially with center-to-center separation $d=12$\,cm and are driven in series by a function
generator and power amplifier (output impedance $0.82\,\Omega$, maximum $6.8$\,V). Typical
operation uses drive frequencies of $12$--$18$\,Hz and currents of $0.6$--$0.8$\,A. At maximum
drive, the field at a coil face is about $80$\,G, and the axial component provides the oscillating
gradient used for trapping.

The trapped object is a donut-shaped NdFeB ring magnet (2.5\,cm OD $\times$ 1.3\,cm ID $\times$
0.25\,cm, 24.1\,g) suspended from a 1.73\,m polyester thread. On one end, the thread is attached to the ceiling with a simple knot, and on the other, to the magnet with a trucker's knot approximately $10$\,cm above the magnet which allows for height adjustment. The
suspension eliminates contact friction while adding only a weak static restoring curvature
characterized by the small-amplitude free-oscillation period $T_p=5.7$\,s
($\omega_p=2\pi/T_p=1.10$\,rad\,s$^{-1}$). This is longer than the geometric pendulum value for
$L=1.73$\,m, which we attribute to weak static magnetostatic coupling to nearby ferromagnetic
coil hardware. We therefore use the measured $\omega_p$ to define the dimensionless static-curvature
parameter $a\equiv 4\omega_p^{2}/\Omega^{2}$. A 0.75\,cm glass rod passes horizontally through the
ring and is used only for alignment of the magnetic axis with the coil axis; during operation,
the magnet remains clear of the rod.

For reproducibility, we calibrated the axial field gradient at the center,
$\left.\partial B_x/\partial x\right|_{x=0}$, using a Hall probe. We measured the axial field at
the midpoint ($B_x(0)\approx 0$) and near the inner face of a coil (at $x\simeq d/2-t/2$, where
$t=2.4$\,cm is the coil thickness), and estimated the midpoint gradient from the ratio
$[B_x(x)-B_x(0)]/x$. We find
$\left.\partial B_x/\partial x\right|_{x=0}\approx (0.24\,\mathrm{T\,m^{-1}\,A^{-1}})\,I$,
allowing others to match the effective midpoint gradient (and thus the micromotion floor) even
with different coil constructions. Under typical conditions, $a=4\omega_p^{2}/\Omega^{2}\sim10^{-3}$,
and the measured secular frequency implies $q_{\rm eff}\ll 1$ (see Sec.~V.B).

For classroom demonstrations, we choose drive frequencies where the secular and micromotion
time scales are visually distinct (12--18\,Hz), and we also show the onset of dynamic instability
at lower frequencies (typically near 6--7\,Hz for our setup). Motion is recorded with a face-on
digital camera at 120\,fps. Pixel--meter calibration uses either the known rod diameter or a
ruler placed in the plane of motion. Positions are extracted with the open-source \texttt{Tracker}
program or an equivalent Python script; data and analysis code are provided with this article.

\section{DEMONSTRATION PROCEDURE}
\label{sec:demo_procedure}

Before powering the coils, ensure that the coil axis, magnet's swing direction, and magnet's face are aligned. The setup is correct when the ring magnet suspends in equilibrium without contact with the glass rod, leaving $1$--$2$\,mm of clearance.
 
\subsection{Trapping the magnet}
Begin the demonstration by showing students the apparatus and the static configuration. Offset the magnet from the center by roughly $1$\,cm to illustrate the weak restoring force from the pendulum and the lack of confinement in the undriven anti-Helmholtz field. Ask students to predict how an oscillating magnetic gradient might stabilize the motion.
Next, apply a sinusoidal current at a frequency below the expected stability threshold. The magnet will strike the rod or coil faces, demonstrating instability. Gradually increase the drive frequency or reduce the drive amplitude until the magnet becomes confined between the coils. Encourage students to discuss how frequency and amplitude affect stability.

\subsection{Exploring stability and motion}
With the magnet trapped, vary the drive frequency and amplitude to show how the secular and micromotion time scales change. Increasing the drive frequency typically improves the confinement margin but reduces the secular oscillation amplitude. Conversely, lowering the frequency below about $6$\,Hz results in loss of confinement. The dual time scales---rapid micromotion superimposed on slower secular motion---are readily visible to the eye.

\subsection{Recording and analysis}

For quantitative analysis, we record the motion with a high-speed camera. Tracking the center of the magnet frame by frame gives its position as a function of time, $x(t)$. To separate the slow and fast motion, we obtain the secular curve $x_{\mathrm{sec}}(t)$ by smoothing $x(t)$ with a Savitzky--Golay filter \cite{savitzky_1964}, which effectively acts as a low-pass filter well below the drive. We then divide the trace into short windows spanning a few drive periods, and in each
window, subtract the local secular value so that the residual has zero average. The micromotion amplitude $A_{\mathrm{mm}}$ in that window is taken from the size of this residual (see Sec.~V.B for details).

Plotting $A_{\mathrm{mm}}$ versus $|x_{\mathrm{sec}}|$, as shown in Fig.~\ref{fig:micro_secular_ratio}, reveals a finite micromotion floor near the center and an even increase with displacement. Over our operating range, the data are well described by the midpoint model $A_{\mathrm{mm}}(X)\simeq A_{\mathrm{floor}}+\beta X^2$ [Eq.~\eqref{eq:Am_floor_beta}], which
quantifies the center ``quietness'' ($A_{\mathrm{floor}}$) and the growth of micromotion toward the turning points ($\beta$). Instructors can display the videos and corresponding plots to help students connect the observed two-timescale dynamics with the underlying ponderomotive restoring mechanism.

\section{RESULTS}
\label{sec:results}

\subsection{Micromotion and secular motion are directly visible}

\begin{figure}[t]
  \centering
  \includegraphics[width=0.48\textwidth]{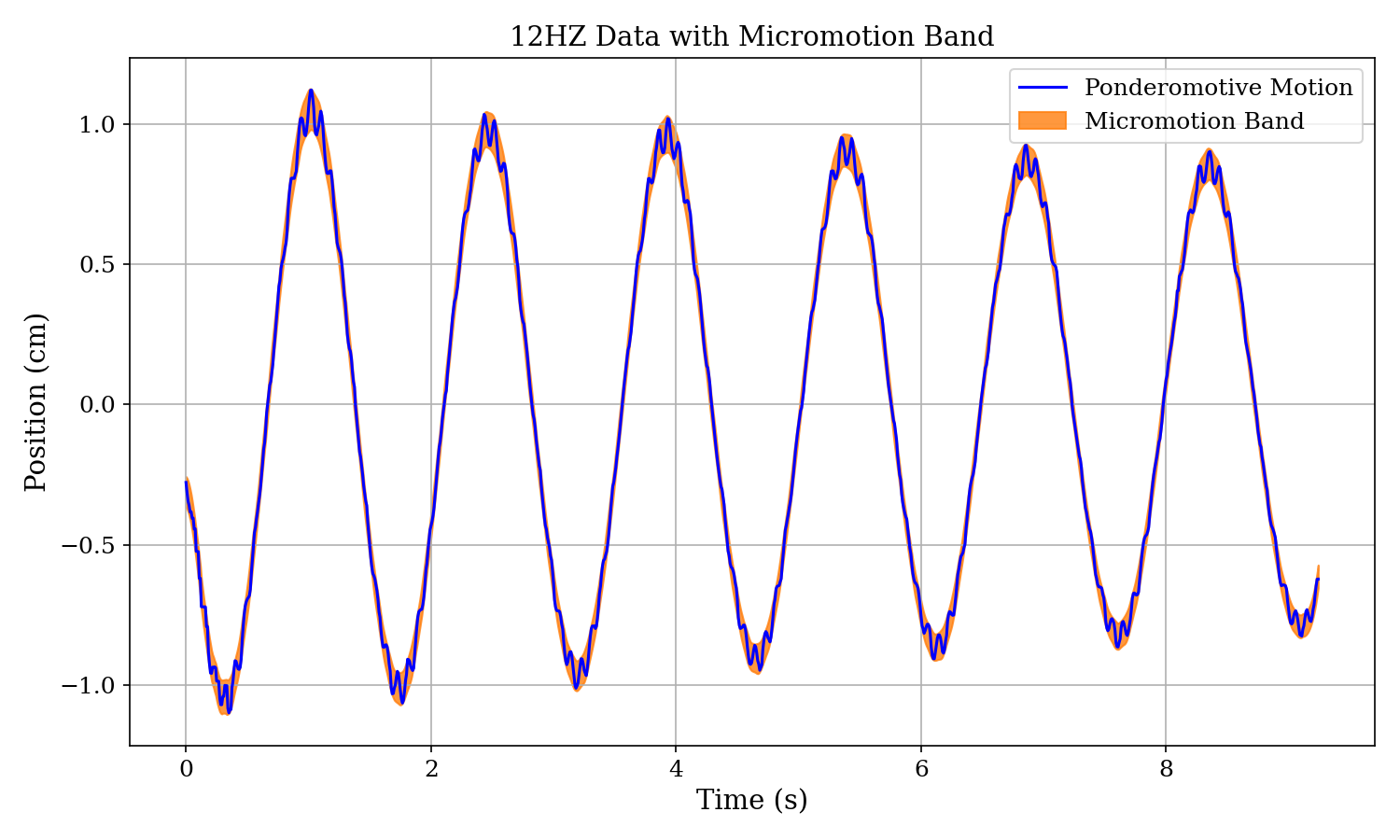}  
  \caption{Position versus time at $12$\,Hz. The blue curve is the measured trajectory; the orange band indicates the micromotion envelope amplitude $A_{\rm mm}(t)$ about the secular motion
$x_{\rm sec}(t)$. Micromotion is maximal at the secular turning points and is minimized near the center, the hallmark of ponderomotive trapping in the small-$q_{eff}$ regime.}
  \label{fig:micromotion12Hz}
\end{figure}

Figure~\ref{fig:micromotion12Hz} shows a representative trajectory at $12$\,Hz. The low-pass
curve (blue) indicates the secular motion $x_{\rm sec}(t)$ while the shaded band indicates the
instantaneous micromotion amplitude $A_{\rm mm}(t)$ about the secular trajectory. The two time
scales are clearly separated: micromotion is smallest near the trap center and largest near the
secular turning points. This qualitative behavior persists across the 12--18\,Hz operating range.

\subsection{Quantifying the timescale separation: $q_{\rm eff}$}
We quantify the operating regime using the timescale separation between secular motion and
micromotion. Specifically, we extract the secular frequency $\omega_{\rm sec}$ and report
\begin{equation}
q_{\rm eff}=2\sqrt{2}\,\frac{\omega_{\rm sec}}{\Omega}
\label{eq:qeff_results}
\end{equation}
(see Sec.~II). We estimate $\omega_{\rm sec}$ in two ways:






\paragraph*{Method 1: frequency ratio.}
We count $N$ drive periods per full secular period, giving
\begin{equation}
\frac{\omega_{\rm sec}}{\Omega}\approx \frac{1}{N},
\qquad
q_{\rm eff}\approx \frac{2\sqrt{2}}{N}.
\label{eq:qeff_from_counts}
\end{equation}
This is a demonstration-friendly method to extract $q_{\rm eff}$. For the $12$\,Hz run, we observe $N\approx 18$, yielding $q_{\rm eff}\approx 0.16$.


\paragraph*{Method 2: secular fit.}
We fit the low-pass secular trajectory to a damped sinusoid,
$x_{\rm sec}(t)\approx X_0 e^{-t/\tau}\cos(\omega_{\rm sec} t+\phi)$,
and extract $\omega_{\rm sec}$. For the 12\,Hz run, the fit yields
$T_{\rm sec}=\,$1.46\,s (so $\omega_{\rm sec}/2\pi=\,0.68$\,Hz), corresponding to
$q_{\rm eff} \approx \,0.16$. This agrees with the cycle-count estimate within uncertainty.

\begin{figure}[t]
  \centering
  \includegraphics[width=0.48\textwidth]{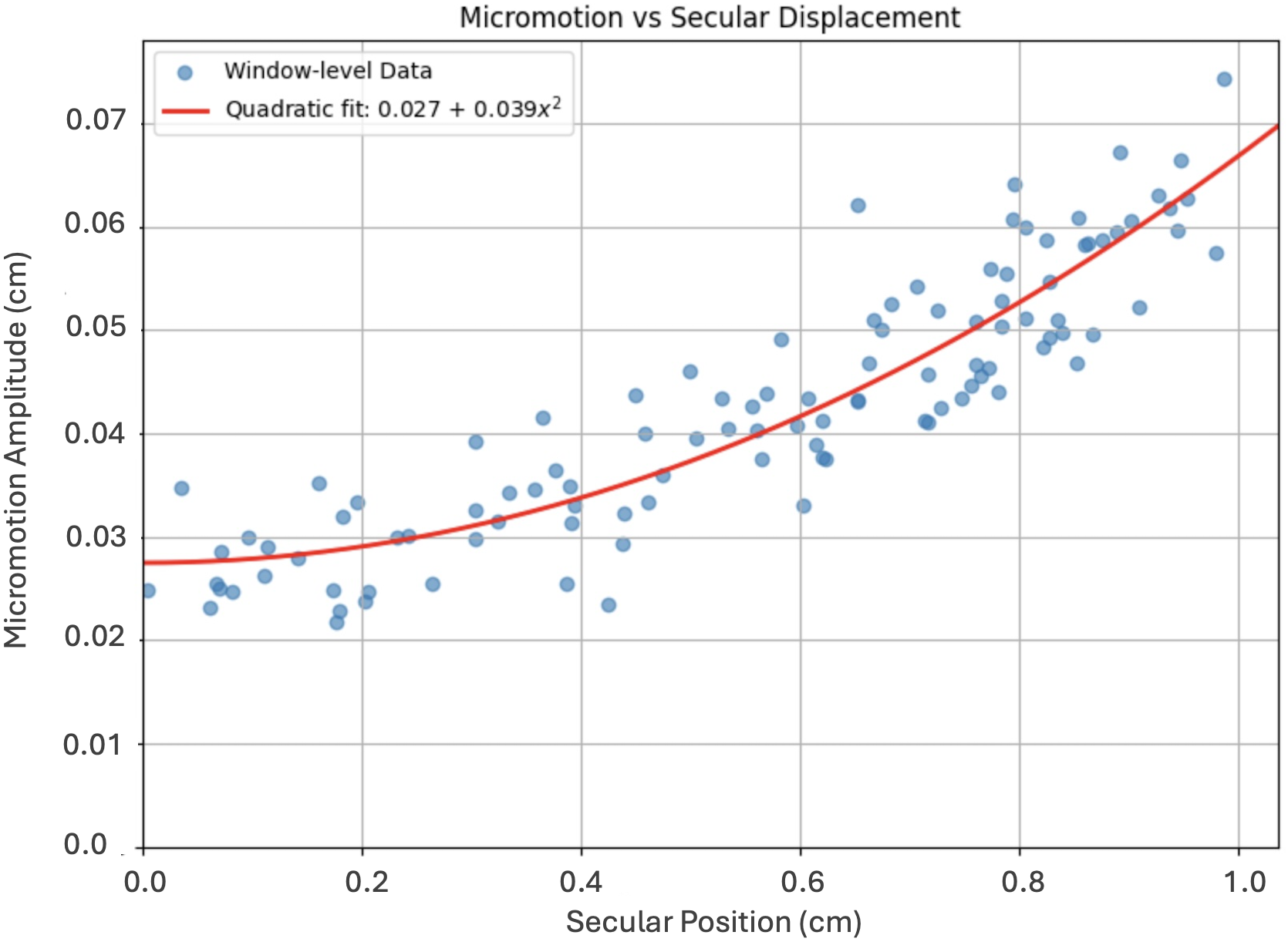}  
  \caption{Micromotion envelope amplitude $A_{\rm mm}$ as a function of secular displacement
$|x_{\rm sec}|$ for a 12\,Hz drive. Each point corresponds to a short time window of two drive
periods. The solid curve is a fit to the midpoint model
$A_{\rm mm}(X)=A_{\rm floor}+\beta X^2$ [Eq.~\eqref{eq:Am_floor_beta}], which captures a finite
micromotion floor near the center and an even increase with displacement toward the turning points.}
  \label{fig:micro_secular_ratio}
\end{figure}

\paragraph*{Micromotion envelope versus secular displacement.}
To quantify the micromotion envelope, we divide the trace into short windows spanning a few
drive periods, subtract the local secular value so the residual has zero mean, and take the
size of the residual as $A_{\rm mm}$ for that window. Figure~\ref{fig:micro_secular_ratio} plots
$A_{\rm mm}$ against $|x_{\rm sec}|$ for a 12\,Hz run. The data show a finite micromotion minimum
near the center and an even increase with displacement toward the turning points. Over our
operating range, the curve is well described by $A_{\rm mm}(X)=A_{\rm floor}+\beta X^2$
[Eq.~\eqref{eq:Am_floor_beta}], from which we extract $A_{\rm floor}$ and $\beta$.

\paragraph*{Small-$a$ check.}
As discussed in Sec.~\ref{sec:exp_setup}, $a\equiv 4\omega_p^2/\Omega^2$ quantifies the weak
static restoring curvature from the suspension. Using $T_p=5.7$\,s
($\omega_p=2\pi/T_p$) gives $a\approx 8.6\times10^{-4}$ at 12\,Hz and $3.8\times10^{-4}$ at 18\,Hz.
These values are $\ll 1$, so the static curvature provides only a small correction to the
high-frequency ponderomotive confinement.

\subsection{Qualitative edge of stability}
This qualitative instability is consistent with the expectation that decreasing the drive frequency increases $q_{\rm eff}\propto \omega_{\rm sec}/\Omega$, eventually spoiling the timescale separation required for robust ponderomotive confinement; we do not attempt a precise mapping of the boundary here.

\section{ROBUSTNESS AND SYSTEMATICS}
\label{sec:systematics}

\begin{itemize}
\item \textbf{Alignment.} Axial and vertical misalignment can introduce unwanted transverse motion and bias the cycle-count estimate of the secular period. We shim the coil mounts and adjust the suspension until out-of-plane motion near the center is minimized.

\item \textbf{Camera sampling and calibration.} Using 120\,fps resolves the drive frequency $\Omega/2\pi$ over 12--18\,Hz with $\sim$7--10 frames per drive cycle. The reported timescale separation, quantified by $q_{\rm eff}=2\sqrt{2}\,\omega_{\rm sec}/\Omega$, is dimensionless and is insensitive to the absolute pixel--meter scale. (Absolute calibration is only required when reporting $A_{\rm mm}$ and $x_{\rm sec}$ in physical units.)

\item \textbf{Secular and envelope estimation.} The secular trajectory $x_{\rm sec}(t)$ is extracted by smoothing $x(t)$ with a Savitzky--Golay filter (30-frame window, low-order polynomial). The micromotion amplitude $A_{\rm mm}$ is computed in short windows spanning a few drive periods by subtracting the local secular value so the residual has zero mean; the size of this residual defines $A_{\rm mm}$ for that window.

\item \textbf{Coil motion and damping.} Air drag and internal losses in the suspension produce a slow decay of the secular amplitude. Reaction forces on the anti-Helmholtz coils can also drive small oscillations of the coil mounts at the drive frequency. Bracing the supports reduces this effect. Such motion can broaden the measured micromotion band and add to the baseline micromotion level, but does not change the extracted $\omega_{\rm sec}$ (and hence $q_{\rm eff}$) within our uncertainties.

\item \textbf{Small positive $a$.} The static curvature from the suspension is small:
$a\equiv 4\omega_p^2/\Omega^2\sim 10^{-3}$ over 12--18\,Hz. It provides only a small correction to $\omega_{\rm sec}$ compared with the ponderomotive contribution and does not alter the qualitative two-timescale dynamics.
\end{itemize}

\section{APPLICATIONS AND EDUCATIONAL VALUE}
\label{sec:education}

Quadrupole ion traps (RF Paul traps) play a central role in modern physics, underpinning
technologies such as optical atomic clocks and trapped-ion quantum computers \cite{ludlow_2015,bruzewicz_2019}. Yet the ponderomotive mechanism behind their confinement is often introduced through rotating-saddle demonstrations, in which a steel ball bearing is stabilized on a mechanically rotating surface \cite{Thompson_2002}. While these demos vividly show that dynamic stability is possible, they also introduce complications that can obscure the underlying two-timescale physics: the drive is rotational rather than a time-varying gradient, the ball both translates and spins as it rolls, and friction can mask how a slow restoring force emerges from a spatial gradient in the fast driving motion \cite{Fan_2017}. Leidenfrost-levitated liquid-nitrogen (LN2) rotating-saddle variants reduce rolling and friction, yielding trajectories closer to ideal point-particle dynamics \cite{Barnett_2025}, though they remain rotating-saddle analogs.

By contrast, the anti-Helmholtz demonstration presented here realizes a one-dimensional
flapping magnetic-field gradient that directly produces a visible micromotion at the drive frequency superimposed on a slower secular oscillation. The suspended ring magnet does not roll, contact friction is negligible, and the motion is confined to a single axis, so the micromotion and secular motion appear in a cleaner and more interpretable way than in rotating-saddle analogs.

The setup is simple enough for an advanced undergraduate laboratory. Students can assemble
and align the apparatus, record trajectories in the 12--18\,Hz range, and use the provided code to extract the secular frequency $\omega_{\rm sec}$ and the dimensionless timescale-separation parameter $q_{\rm eff}=2\sqrt{2}\,\omega_{\rm sec}/\Omega$ by independent methods. In lecture, short demonstrations can highlight key features---for example, that micromotion is minimized near the center and largest near the turning points, or that lowering the drive frequency can lead to loss of confinement near 6--7\,Hz in our setup. The demonstration can therefore be adapted to interactive lectures, guided lab activities, or student-driven projects in courses on classical mechanics, electromagnetism, or modern/AMO physics.

\section{CONCLUSION}
\label{sec:conclusion}

We presented a compact mechanical analog of an RF Paul trap that makes the ponderomotive
mechanism directly visible. Unlike rotating-saddle demonstrations, which illustrate dynamic stability through geometric rotation and rolling motion, the anti-Helmholtz flapping trap realizes a one-dimensional time-varying field gradient with negligible contact friction and a clearly interpretable trajectory.

In the high-frequency regime the motion separates into a slow secular oscillation and a rapid micromotion at the drive frequency. Consistent with the anti-Helmholtz symmetry about the midpoint, the micromotion is minimized at the center and increases toward the secular turning points; the measured envelope is well described by the midpoint model $A_{\rm mm}(X)\simeq A_{\rm floor}+\beta X^2$. Despite this nonlinear envelope, high-frequency averaging yields a harmonic ponderomotive potential to leading order near the center, explaining the sinusoidal secular motion.

Using video tracking and straightforward data analysis, we extract the secular frequency
$\omega_{\rm sec}$ and quantify the operating regime via the dimensionless timescale-separation parameter $q_{\rm eff}=2\sqrt{2}\,\omega_{\rm sec}/\Omega$. We also observe loss of confinement as the drive frequency is lowered toward 6--7\,Hz in our setup. Together with the materials list, field-gradient calibration, and analysis workflow provided with this article, these results make the demonstration practical for lecture and laboratory use and provide students with a clear, intuitive picture of ponderomotive confinement in RF Paul traps.

\section*{ACKNOWLEDGMENTS}
We are grateful for helpful discussions with Laurel Barnett and Aidan Carey. We gratefully acknowledge support from the Faculty of Arts and Sciences at Harvard University. ChatGPT was used to refine this manuscript.

\section*{AUTHOR DECLARATIONS}
The authors have no conflicts to disclose.

\section*{APPENDIX}

\appendix
\section{Averaging derivation of the ponderomotive potential}
\label{appendix:ponderomotive_derivation}

Consider the one-dimensional equation of motion for a particle
of mass $m$ subject to a rapidly oscillating drive force with slowly varying amplitude,
\begin{equation}
m\ddot{x} = F_{\rm slow}(x) + F_0(x)\cos(\Omega t),
\label{eq:eom_appendix}
\end{equation}
where $F_{\rm slow}(x)$ represents any weak, slowly varying force (e.g., static suspension
curvature) and $F_0(x)$ varies slowly with position but oscillates rapidly in time.

We separate the total motion into slow and fast components, $x(t)=X(t)+\xi(t)$, where $X(t)$
is the secular coordinate and $\xi(t)$ is the micromotion \cite{dehmelt_1968}. Expanding about $X$ and keeping
leading terms gives
\begin{equation}
m\ddot{X} + m\ddot{\xi}
\simeq
F_{\rm slow}(X) + F_0(X)\cos(\Omega t)
+ \xi\,\frac{dF_0}{dX}\cos(\Omega t),
\end{equation}
where we have neglected higher-order terms in $\xi$ and any terms involving derivatives of
$F_{\rm slow}$, which are small in the high-frequency regime.

The fast motion is dominated by the oscillatory drive and satisfies, to leading order,
\begin{equation}
m\ddot{\xi} \simeq F_0(X)\cos(\Omega t),
\end{equation}
whose solution is
\begin{equation}
\xi(t) \simeq -\frac{F_0(X)}{m\Omega^2}\cos(\Omega t).
\label{eq:xi_appendix}
\end{equation}
Substituting Eq.~\eqref{eq:xi_appendix} back into the full equation and averaging over one
drive period (so that $\langle\cos(\Omega t)\rangle=0$ and $\langle\cos^2(\Omega t)\rangle=\tfrac{1}{2}$)
yields the secular equation
\begin{equation}
m\ddot{X} \;=\; F_{\rm slow}(X) \;-\;\frac{1}{4m\Omega^2}\frac{d}{dX}\!\left[F_0^2(X)\right].
\label{eq:secular_appendix}
\end{equation}
This defines an effective (ponderomotive) potential energy,
\begin{equation}
U_{\rm pond}(X)=\frac{F_0^2(X)}{4m\Omega^2},
\label{eq:Upond_appendix}
\end{equation}
so that the averaged force can be written $m\ddot{X}=F_{\rm slow}(X)-dU_{\rm pond}/dX$.

Equation~\eqref{eq:Upond_appendix} also shows that the cycle-averaged kinetic energy of the
micromotion,
\begin{equation}
\langle T_{\rm mm}\rangle
=\left\langle \tfrac{1}{2}m\dot{\xi}^{2}\right\rangle
=\frac{F_0^2(X)}{4m\Omega^2},
\end{equation}
acts as an effective potential energy whose spatial gradient produces the slow restoring
force responsible for the observed secular motion.

\paragraph{Application to the anti-Helmholtz midpoint expansion.}
In the present work, the oscillatory drive amplitude takes the form
$F_0(X)=F_c+\alpha X^2+O(X^4)$ near the midpoint (main text, Sec.~II), so
\begin{equation}
U_{\rm pond}(X)=\frac{(F_c+\alpha X^2)^2}{4m\Omega^2}
\simeq \text{const}+\frac{F_c\alpha}{2m\Omega^2}X^2+O(X^4).
\end{equation}
Thus, even though the micromotion envelope is not linear in $X$, the resulting time-averaged
potential is harmonic to leading order near $X=0$, consistent with the approximately sinusoidal
secular motion observed in the trajectories.

\section{Materials and cost}
Table I summarizes the primary components required for the 1-D ponderomotive demonstration, including vendors, estimated costs, and relevant comments. Although the table may appear later due to float placement, it is part of this subsection.

\squeezetable

\begin{table*}[t]
\centering
\squeezetable
\caption{Materials and approximate costs of demonstration setup.}
\begin{ruledtabular}
\begin{tabular}{cllll}
Item & Name & Vendor and Model & Price (USD) & Comment \\ \hline
1 & Anti-Helmholtz coils & --- & --- & A pair of parallel, coaxial coils on a stable track\\
2 & Ring magnet & K\&J Magnets: 1 x 1/2 x 1/8 Inch Neodymium Rare Earth Ring Magnet & 3.19 & ---\\
3 & String & Thin polyester thread & 2.50 & Price per 100 yds; 0.002 g/cm \\
4 & Rod & Glass rod & 8.00 & Price per 12 inches; ~0.3-inch outer diameter \\
5 & Signal generator & Pasco Scientific PI-9587C Digital Function Generator-Amplifier & 250.00 & Resale on Ebay \\
6 & Electric wires & Alligator clip wires & 1.00 & --- \\
\end{tabular}
\end{ruledtabular}
\end{table*}

\squeezetable

\section{Calibrated field-gradient–versus–current data}

For a current of $0.7\,\mathrm{A}$, we detected a field of $80\,\mathrm{G}$ on one coil.
We estimated the gradient by approximating it to be linear near the center, which gave a field gradient of
$0.17\,\mathrm{T\,m^{-1}}$.


\bibliography{references} 

\end{document}